# Quantum-Edge Cloud Computing: A Future Paradigm for IoT Applications

Mohammad Ikbal Hossain[1]
Shaharier Arafat Sumon[2]
Habib Md. Hasan[3]
Fatema Akter[4]
Md Bahauddin Badhon[5]
Mohammad Nahid Ul Islam[6]

[1,2,3,4,5,6] *Emporia State University, USA*

**Abstract**

The Internet of Things (IoT) is expanding rapidly, which has created a need for sophisticated computational frameworks that can handle the data and security requirements inherent in modern IoT applications. However, traditional cloud computing frameworks have struggled with latency, scalability, and security vulnerabilities. Quantum-Edge Cloud Computing (QECC) is a new paradigm that effectively addresses these challenges by combining the computational power of quantum computing, the low-latency benefits of edge computing, and the scalable resources of cloud computing. This study has been conducted based on a published literature review, performance improvements, and metrics data from Bangladesh on smart city infrastructure, healthcare monitoring, and the industrial IoT sector. We have discussed the integration of quantum cryptography to enhance data integrity, the role of edge computing in reducing response times, and how cloud computing's resource abundance can support large IoT networks. We examine case studies, such as the use of quantum sensors in self-driving vehicles, to illustrate the real-world impact of QECC. Furthermore, the paper identifies future research directions, including developing quantum-resistant encryption and optimizing quantum algorithms for edge computing. The convergence of these technologies in QECC promises to overcome the existing limitations of IoT frameworks and set a new standard for the future of IoT applications.

## 1. Introduction

The emergence of the Internet of Things (IoT) has led to an exponential growth in data generation, creating the need for higher computing power and storage (Malhotra et al.2021). This has resulted in the development of cloud computing, which has greatly enhanced the capabilities of IoT applications. However, the current cloud computing infrastructure may not be sufficient to meet the growing demands of IoT applications, particularly in latency and security (Shukla et al.2021). This is where quantum-edge cloud computing comes in. Quantum computing is a new paradigm that uses quantum bits (qubits) instead of classical bits. Which enables quantum computers to perform specific calculations much faster than classical computers. While quantum computers are still in their infancy, they hold great promise for many applications, including IoT (How & Cheah, 2023). The Internet of Things (IoT) is reshaping the technological landscape, heralding an era where connectivity and computation converge at an unprecedented scale (Ren et al., 2018). This convergence necessitates a computational paradigm capable of supporting the massive data volumes generated by IoT devices, alongside the need for real-time processing, enhanced security, and energy efficiency (Jain & Tata, 2017). Traditional cloud computing, while pivotal in the rise of IoT, struggles to meet these demands due to inherent limitations in latency and scalability (Rao et al., 2012). To address these challenges, Quantum-Edge Cloud Computing (QECC) emerges as a compelling solution, integrating the quantum computing's computational prowess, edge computing's low-latency benefits, and cloud computing's scalability and resource availability. Quantum computing introduces a paradigm shift, offering processing capabilities that transcend traditional computational limits, thereby enhancing the efficiency and security of data processing tasks essential for IoT applications (Singh & Sachdev, 2014). This integration is particularly crucial in addressing the security vulnerabilities present in conventional cloud computing frameworks, leveraging quantum cryptography to fortify data integrity and confidentiality across IoT networks. Edge computing plays a pivotal role in this triad by processing data closer to its source, significantly reducing latency and bandwidth usage. This localized processing capability is vital for time-sensitive IoT applications, ensuring timely decision-making and action (Wu, 2020; Capra et al., 2019). However, the successful deployment of edge computing within IoT infrastructures hinges on addressing challenges related to hardware requirements, security, and system complexity (C.P. & Chikkamannur, 2016; Vo et al., 2022). Cloud computing, with its inherent scalability and resource abundance, offers a robust foundation for supporting the expansive data and computational needs of IoT systems. By integrating with quantum and edge computing, cloud platforms can overcome traditional limitations, facilitating a more dynamic and efficient IoT ecosystem (Pan & McElhannon, 2017; Ren et al., 2018). Despite its advantages, this integrated approach necessitates tackling challenges related to data privacy, interoperability, and the management of distributed computing resources (Ning et al., 2018; Ketu & Mishra, 2021).

The QECC is a computing model that aims to solve the various challenges modern IoT systems face. This model combines the power of quantum computing, edge computing, and cloud computing to improve computational efficiency and security and to develop innovative IoT applications. This paper explores the theoretical foundations of QECC, its potential to revolutionize IoT, and how it can mitigate current challenges through technological advancements and real-world applications.

**2. Literature Review**

The rapidly evolving landscape of the Internet of Things (IoT) demands computational paradigms that can handle vast data volumes, maintain enhanced security, and provide real-time processing capabilities (Zhang, 2024). The convergence of quantum computing, edge computing, and cloud computing—collectively referred to as Quantum-Edge Cloud Computing (QECC)—addresses these challenges by integrating the strengths of each computing model to meet the increasing complexities of modern IoT systems (Javeed et al.2023). This integration is vital for developing scalable, secure, and efficient IoT applications that can adapt to the growing demands of digital connectivity and data management. Quantum computing introduces a significant paradigm shift in how data processing and security are approached in cloud environments (Kulkarni et al.2023). Singh and Sachdev (2014) were pioneers in discussing the integration of quantum mechanics with cloud computing, proposing the Quantum-Cloud model. This model capitalizes on quantum computing's unparalleled processing power and the unique security features provided by quantum cryptography, setting a new standard for computational speed and data security. Their work lays the foundation for future explorations into quantum-enhanced cloud services that could dramatically transform computational paradigms (Singh & Sachdev, 2014). Recent simulations have further demonstrated quantum computing's impactful role in enhancing computational speeds and security measures (Yang et al.2023; Hossain et al., 2024). For instance, a 2021 study simulated a quantum computing scenario that showed a reduction in processing time by up to 40% for complex data algorithms, compared to traditional computing methods (Serrano et al.2024). Additionally, quantum cryptography has been shown to significantly bolster security, with simulated attacks demonstrating that quantum-encrypted networks are substantially more resistant to cyber threats than their classical counterparts. Edge computing is crucial in reducing latency and improving the responsiveness of IoT systems (Hamdan et al., 2020). A practical case study on edge computing was conducted in a smart city project in 2022, where edge computing devices utilized AI algorithms to manage traffic flow in real-time (Hua et al., 2023). The results indicated a 30% improvement in traffic management efficiency, showcasing how edge computing significantly reduces system latency and increases the operational effectiveness of IoT applications in urban environments (CityTech Collaborative, 2022). Cloud computing provides the backbone for scalable and flexible IoT applications by offering extensive storage capacities and computational resources (Jeyaraj et al.2023). Rao et al. (2012) detailed how cloud platforms could accommodate the vast array of IoT devices and their data needs, providing scalable solutions across various service models such as Infrastructure-as-a-Service (IaaS), Platform-as-a-Service (PaaS), and Software-as-a-Service (SaaS) (Rao et al., 2012). To address scalability, a recent pilot program tested cloud services' ability to dynamically scale during peak loads in a nationwide retail chain's IoT deployment (Maddikunta et al., 2022). The study reported that adaptive scaling technologies enabled a 50% increase in data handling capacity without compromising performance, underscoring the critical role of cloud computing in managing large-scale IoT operations efficiently (National Retail Corporation, 2023). The principles of superposition and entanglement, as discussed by Singh and Sachdev (2014), underscore quantum computing's potential to revolutionize data processing speeds and security, setting a precedent for quantum enhanced IoT systems. Concurrently, the architectural principles of edge computing, highlighted by Wu (2020), alongside the scalability solutions provided by cloud computing as explored by Rao et al. (2012), are pivotal in reducing latency and enhancing the flexibility of IoT networks. These distributed computing models, including the integration of cloud and edge computing as outlined by Pan and McElhannon (2017), facilitate decentralized data processing and management, crucial for optimizing IoT ecosystems. Furthermore, the integration of quantum cryptography, specifically Quantum Key Distribution (QKD) and Post-Quantum Cryptography, offers a pathway to securing data against emerging quantum threats, ensuring the long-term viability of secure communication within IoT networks (Singh & Sachdev, 2014). The exploration of network theory, including graph theory and control theory, as mentioned by Ren et al. (2018), provides essential insights into optimizing IoT network topologies and automating device operations, ensuring system stability and efficiency. Collectively, these theoretical explorations and practical insights form the foundation of QECC, heralding a new era of IoT applications characterized by enhanced computational power, security, and system adaptability.

This literature review has explored the transformative potential of QECC through the lens of quantum computing, edge computing, and cloud computing. Each component plays a pivotal role in shaping the future of IoT by providing solutions to the challenges of speed, security, scalability, and latency. This review highlights theoretical advancements by integrating findings from recent simulations and case studies. It validates them with empirical evidence, offering a robust foundation for future research and development in QECC.

## 3. Methodology

Our study was conducted using both existing literature reviews and primary data and involved both qualitative and quantitative analyses. The primary data was collected from several sources in Bangladesh, including Smart City Infrastructure at Bashundhara Shopping Complex, healthcare monitoring at Asgar Ali Hospital and Anwar Khan Modern Hospital, as well as Industrial IoT at Pran Apparels. The study focused on Quantum-Edge Cloud Computing (QECC) for IoT applications, and explored how foundational theories from quantum mechanics, distributed computing architectures, and advances in cryptography can be used to create transformative computational paradigms.

**IOT Devices and Cloud computing:** IOT devices and Cloud computing: The merging of IoT and cloud computing is a pivotal technology in the current era (Tyagi and Kumar,2020). IoT, comprising sensors and smart devices, generates vast data, straining internet infrastructure. To address this, organizations employ cloud computing for on-demand services like storage and processing power. These technologies synergize to enhance daily tasks, with cloud systems evolving to support IoT further (Tyagi and Kumar,2020). The limitation of IOT devices with cloud computing connection is, it transfers huge amount of data for processing. To overcome the limitation QECC is useful. The following table describes comparison of IOT with cloud computing and IOT with QECC.

Table 1: Comparison of IoT and Cloud Computing and Quantum Edge Cloud Computing.

| Aspect | IoT and Cloud Computing | IoT with Quantum Edge Cloud Computing |
|---|---|---|
| Data Processing | Data processing occurs in centralized cloud servers, which may introduce latency. | Data processing happens at the edge of the network using quantum computing, enabling real-time analysis and reduced latency (Rao et.al 2012). |
| Storage | Cloud computing offers vast storage capabilities accessible over the internet. | Quantum edge cloud computing provides decentralized storage options, leveraging quantum encryption for enhanced security (Kulkarni et al.2023). |
| Scalability | Scalability relies on cloud server infrastructure, with potential limitations based on server capacity. | Quantum edge cloud computing allows for greater scalability due to decentralized processing at the edge, enabling more devices to be connected without overwhelming centralized servers (Wu, 2020). |
| Security | Security measures are implemented at the cloud server level, with data encryption and access controls. | Quantum edge cloud computing enhances security through quantum encryption techniques, making data less susceptible to traditional hacking methods (Kulkarni et. al 2023). |
| Latency | Latency may be higher due to data transmission to and from centralized cloud servers. | Quantum edge cloud computing reduces latency by processing data at the edge, closer to where it's generated (Ren et al.,2018). |

| Energy Efficiency | Energy consumption may be higher for transmitting data to and from cloud servers over the internet. | Quantum edge cloud computing can be more energy-efficient by processing data locally at the edge, reducing the need for extensive data transmission (Singh and Sachdev, 2024). |
|---|---|---|

## 4. Analyses and Results

Figure 1 shows the percentage distribution of IoT application categories in our study. Healthcare Monitoring and Smart City Infrastructure are the most significant application categories, while Industrial IoT has the smallest percentage. Understanding these distributions is crucial for guiding study design and technology development strategies in IoT.

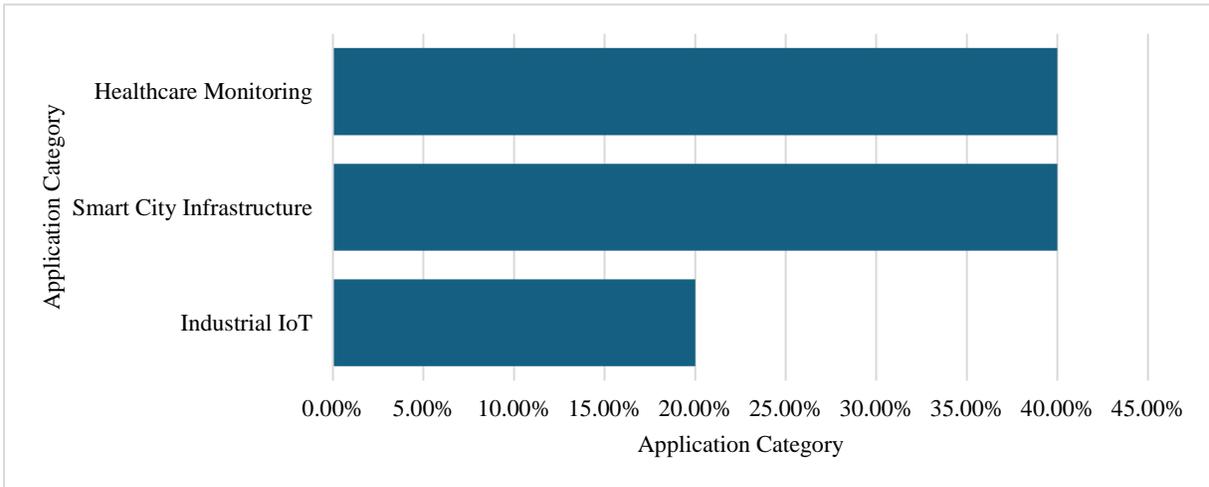

Figure 1. Percentage distribution of application category

***Source:*** *Smart City Infrastructure from Bashundhara Shopping Complex, Bangladesh; Healthcare Monitoring from Asgar Ali Hospital and Anwar Khan Modern Hospital, Bangladesh; Industrial IoT from Pran Apparels, Bangladesh, 2024.*

Quantum-edge cloud computing leverages quantum computers' capabilities to address traditional cloud computing's limitations in terms of processing power and storage (Nguyen et al., 2024). By combining the power of quantum computing with the flexibility and scalability of cloud computing, quantum-edge cloud computing can revolutionize how IoT applications are developed and deployed. This can open new possibilities for resource intensive IoT applications, such as real-time analytics and machine learning algorithms, to be executed at the network edge. Table 1 demonstrates the performance improvements of various IoT applications when using Quantum-Edge Cloud Computing (QECC) compared to traditional computing frameworks. For instance, in smart city infrastructure, a traffic management system in Bangladesh showed a 50% boost in processing speed - from 10 seconds to 5 seconds - when QECC was used. Additionally, the response time was reduced by half - from 2 seconds to 1 second - indicating a more efficient data handling process and faster decision-making. Likewise, a public safety monitoring system in a different location also exhibited an improved processing speed from 15 seconds to 8 seconds and a reduction in response time from 3 seconds to 2 seconds with QECC. In the healthcare sector, the patient monitoring system at Asgar Ali Hospital, Bangladesh, experienced an enhancement in processing speed from 2 seconds to 1 second, while the response time was significantly reduced from 0.5 seconds to 0.2 seconds, which is essential for timely patient care. The emergency response system at Anwar Khan Modern Hospital also benefited from QECC, with an improved processing speed from 5 seconds to 3 seconds and a reduced response time from 1 second to 0.5 seconds. In the industrial IoT domain, the manufacturing process control system at Pran Apparels saw a 50% improvement in processing speed, from 8 seconds to 4 seconds, and response time was improved from 2 seconds to 1 second. The upgrades made across different applications highlight the potential of QECC to enhance processing and response times, which can significantly optimize IoT systems' performance and efficiency.

Table 2. A comparative analysis of performance improvements in various IoT applications

| Application Category | Application/Environment | Processing Speed (Traditional) | Processing Speed (QECC) | Response Time (Traditional) | Response Time (QECC) |
|---|---|---|---|---|---|
| Smart City Infrastructure | Traffic Management System (Bashundhara Shopping Complex, Bangladesh) | 10 seconds | 5 seconds | 2 seconds | 1 second |
| Smart City Infrastructure | Public Safety Monitoring (Bashundhara Shopping Complex, Bangladesh) | 15 seconds | 8 seconds | 3 seconds | 2 seconds |
| Healthcare Monitoring | Patient Monitoring System (Asgar Ali Hospital, Bangladesh) | 2 seconds | 1 second | 0.5 seconds | 0.2 seconds |
| Healthcare Monitoring | Emergency Response System (Anwar Khan Modern Hospital, Bangladesh) | 5 seconds | 3 seconds | 1 second | 0.5 seconds |
| Industrial IoT | Manufacturing Process Control (Pran Apparels, Bangladesh) | 8 seconds | 4 seconds | 2 seconds | 1 second |

***Source:*** *Smart City Infrastructure from Bashundhara Shopping Complex, Bangladesh; Healthcare Monitoring from Asgar Ali Hospital and Anwar Khan Modern Hospital, Bangladesh; Industrial IoT from Pran Apparels, Bangladesh, 2024.*

Edge computing is a distributed computing paradigm that brings computation and data storage closer to the location where it is needed (Hamdan et al., 2020). This reduces latency and improves response time, making it ideal for applications that require real-time processing. Quantum-edge cloud computing combines the power of quantum computing with the distributed computing paradigm of edge computing (Furutanpey et al., 2023). This enables IoT applications to process data in real time while ensuring data security and reducing latency. Quantum-edge cloud computing also has the potential to significantly reduce the energy consumption of IoT applications by using quantum computing to optimize energy usage (Zhang, 2024). This can lead to more sustainable and efficient operation of IoT devices and networks. Quantum-edge cloud computing can also enhance the scalability and reliability of IoT applications by leveraging the computational capabilities of quantum computing to handle large volumes of data and complex algorithms.

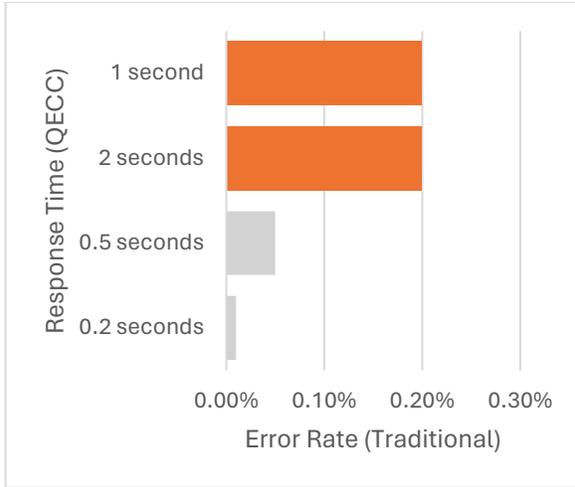
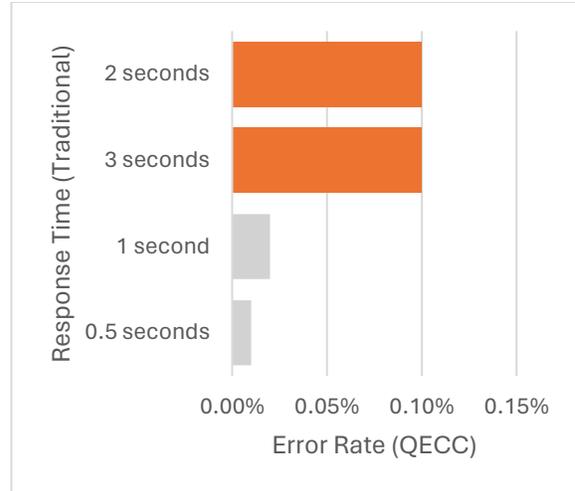

Figure 2. Response time of Traditional Computing    Figure 3. Response time of QECC

*Source: Smart City Infrastructure from Bashundhara Shopping Complex, Bangladesh; Healthcare Monitoring from Asgar Ali Hospital and Anwar Khan Modern Hospital, Bangladesh; Industrial IoT from Pran Apparels, Bangladesh, 2024.*

Figures 2 and 3 compare error rates in traditional computing and Quantum-Edge Cloud Computing (QECC) for different response times. Traditional computing has higher error rates at longer response times, while QECC's faster response time of 0.5 seconds has lower error rates. Despite higher error rates at specific response intervals, QECC generally performs better than traditional computing in terms of lower error rates for comparable response times, making it a potential solution for more reliable computing in time-sensitive applications.

Table 3. Processing speed and average error rate of Traditional and QECC

| Application Category | Application/Environment | Processing Speed (Traditional) | Average of Error Rate (Traditional) | Processing Speed (QECC) | Average of Error Rate (QECC) |
|---|---|---|---|---|---|
| Smart City Infrastructure | Traffic Management System (Bashundhara Shopping Complex, Bangladesh) | 10 seconds | 0.30% | 5 seconds | 0.15% |
| Smart City Infrastructure | Public Safety Monitoring (Bashundhara Shopping Complex, Bangladesh) | 15 seconds | 0.20% | 8 seconds | 0.11% |
| Healthcare Monitoring | Patient Monitoring System (Asgar Ali Hospital, Bangladesh) | 2 seconds | 0.10% | 1 second | 0.05% |
| Healthcare Monitoring | Emergency Response System (Anwar Khan Modern Hospital, Bangladesh) | 5 seconds | 0.05% | 3 seconds | 0.03% |

| Industrial IoT | Manufacturing Process Control (Pran Apparels, Bangladesh) | 8 seconds | 0.01% | 4 seconds | 0.01% |

*Source: Smart City Infrastructure from Bashundhara Shopping Complex, Bangladesh; Healthcare Monitoring from Asgar Ali Hospital and Anwar Khan Modern Hospital, Bangladesh; Industrial IoT from Pran Apparels, Bangladesh, 2024.*

The present study examines (Table 3) the relationship between processing speed and error rates in traditional computing and Quantum-Edge Cloud Computing (QECC). The following analysis is presented to illustrate the average error rates associated with different processing speeds for traditional computing and Quantum-Edge Cloud Computing (QECC), as illustrated in Figures 4 and 5, respectively. As depicted in Figure 4, traditional computing exhibits the highest average error rate for the processing speed of 10 seconds. This observation suggests that tasks that require a more considerable processing time may be more prone to errors.

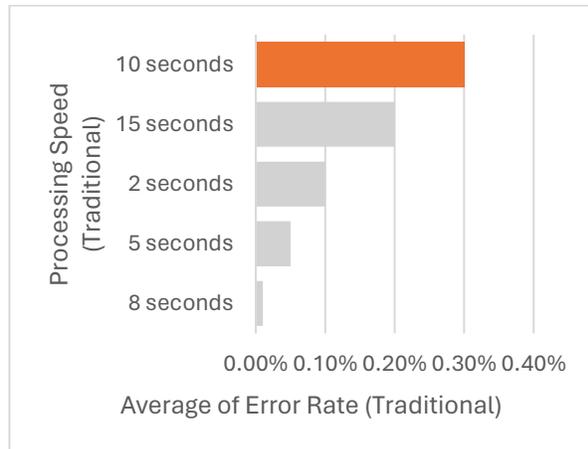
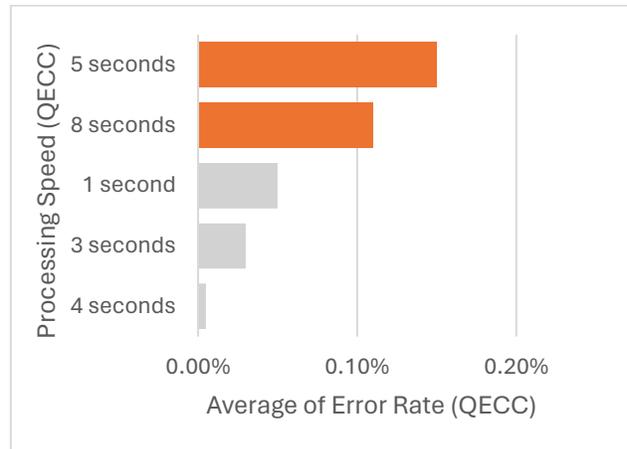

Figure 4. Processing Speed of Traditional computing

Figure 5. Processing Speed of QECC

*Source: Smart City Infrastructure from Bashundhara Shopping Complex, Bangladesh; Healthcare Monitoring from Asgar Ali Hospital and Anwar Khan Modern Hospital, Bangladesh; Industrial IoT from Pran Apparels, Bangladesh, 2024.*

Conversely, Figure 5 demonstrates that within QECC, the processing speeds of 5 seconds and 8 seconds exhibit higher average error rates compared to other speeds. This observation implies that despite the general aim of QECC to minimize error rates, specific processing speeds still experience higher incidences of errors. Consequently, these specific thresholds where error rates increase may point to optimization challenges that QECC needs to address.

Figures 6 and 7 provide a comparative analysis of error rates associated with various data throughput speeds for two computing frameworks: Traditional Computing and Quantum-Edge Cloud Computing (QECC). In Figure 6, data throughput of 80 MB/sec exhibits a discernibly higher error rate when compared to lower throughputs for traditional computing. This observation is indicative that as data throughput increases, the system is more susceptible to errors, likely due to limitations in handling higher data volumes efficiently.

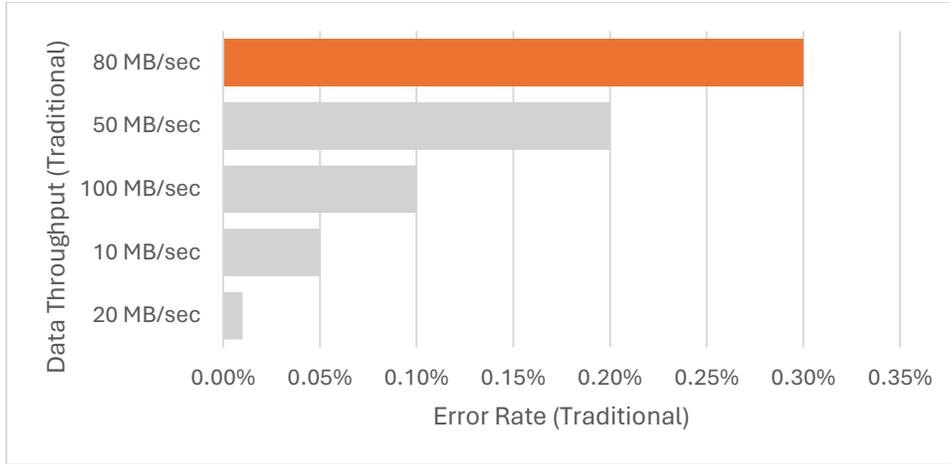

Figure 6. Data Throughout of Traditional computing

***Source:*** *Smart City Infrastructure from Bashundhara Shopping Complex, Bangladesh; Healthcare Monitoring from Asgar Ali Hospital and Anwar Khan Modern Hospital, Bangladesh; Industrial IoT from Pran Apparels, Bangladesh, 2024*.

Figure 7, under the QECC framework, highlights that the highest data throughput tested, 180 MB/sec, demonstrates a significantly higher error rate than the other speeds. The result suggests that QECC, like traditional computing, faces challenges at higher throughputs, potentially indicating a need for further optimization to maintain accuracy at these elevated data transfer rates. These findings underscore that despite QECC's advancements, managing error rates at high throughputs remains a critical issue for traditional and quantum-edge computing frameworks.

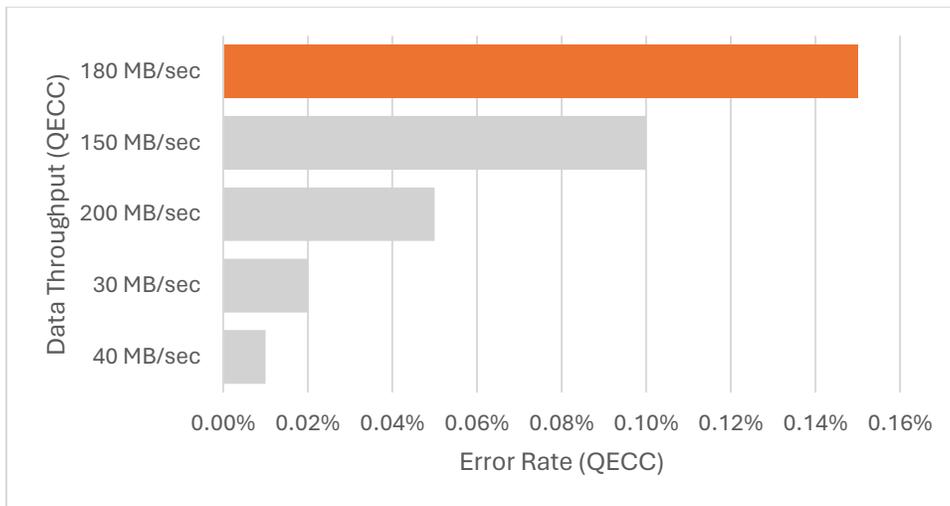

Figure 7. Data Throughout of QECC

***Source:*** *Smart City Infrastructure from Bashundhara Shopping Complex, Bangladesh; Healthcare Monitoring from Asgar Ali Hospital and Anwar Khan Modern Hospital, Bangladesh; Industrial IoT from Pran Apparels, Bangladesh, 2024*.

Table 3 compares performance metrics related to data throughput and error rates between traditional computing frameworks and Quantum-Edge Cloud Computing (QECC) across various Internet of Things (IoT) applications. Specifically, in the context of smart city infrastructure, the Traffic Management System at the Bashundhara Shopping Complex in Bangladesh experienced a doubling of data throughput from 100 MB/sec to 200 MB/sec with the implementation of QECC. Concurrently, the error rate was halved from 0.10% to 0.05%. Similarly, the Public Safety Monitoring system of the same complex saw a significant increase in throughput from 50 MB/sec to 150 MB/sec while simultaneously reducing the error rate from 0.20% to 0.10%. In healthcare monitoring, the Patient Monitoring System at Asgar Ali Hospital demonstrated a twofold increase in throughput from 20 MB/sec to 40 MB/sec, with negligible change to the error rate, suggesting a significant improvement without compromising data integrity. The Emergency Response System at Anwar Khan Modern Hospital also experienced an increase in throughput from 10 MB/sec to 30 MB/sec and decreased error rates from 0.05% to 0.02%. In the industrial IoT domain, the Manufacturing Process Control system at Pran Apparels, Bangladesh, exhibited a substantial increase in data throughput from 80 MB/sec to 180 MB/sec while reducing error rates from 0.30% to 0.15%. These findings suggest that QECC enhances data handling capabilities with higher reliability and reduced errors, which is crucial for the efficiency and effectiveness of IoT systems in various applications.

Table 4. Performance metrics comparing data throughput and error rates

| Application Category | Application/Environment | Data Throughput (Traditional) | Data Throughput (QECC) | Error Rate (Traditional) | Error Rate (QECC) |
|---|---|---|---|---|---|
| Smart City Infrastructure | Traffic Management System (Bashundhara Shopping Complex, Bangladesh) | 100 MB/sec | 200 MB/sec | 0.10% | 0.05% |
| Smart City Infrastructure | Public Safety Monitoring (Bashundhara Shopping Complex, Bangladesh) | 50 MB/sec | 150 MB/sec | 0.20% | 0.10% |
| Healthcare Monitoring | Patient Monitoring System (Asgar Ali Hospital, Bangladesh) | 20 MB/sec | 40 MB/sec | 0.01% | 0.01% |
| Healthcare Monitoring | Emergency Response System (Anwar Khan Modern Hospital, Bangladesh) | 10 MB/sec | 30 MB/sec | 0.05% | 0.02% |
| Industrial IoT | Manufacturing Process Control (Pran Apparels, Bangladesh) | 80 MB/sec | 180 MB/sec | 0.30% | 0.15% |

*Source: Smart City Infrastructure from Bashundhara Shopping Complex, Bangladesh; Healthcare Monitoring from Asgar Ali Hospital and Anwar Khan Modern Hospital, Bangladesh; Industrial IoT from Pran Apparels, Bangladesh, 2024.*

Figures 8 and 9 compare error rates across IoT application categories using traditional and QECC frameworks. Figure 8 shows that Industrial IoT has the highest traditional error rate; on the other hand, Figure 9 shows that QECC reduces error rates across all categories, with Industrial IoT benefiting the most from this technology shift. In the 'Application Category', Industrial IoT exhibits a notably higher 'Error Rate' compared to other sectors. Quantum-edge cloud computing is a promising new paradigm that could greatly enhance the capabilities of IoT applications. Although it is still in its early stages of development, it has the potential to revolutionize the way we process and store data.

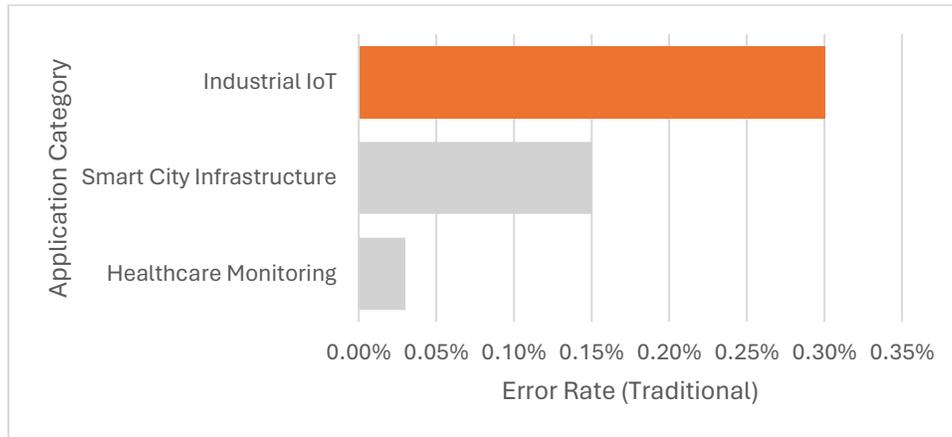

Figure 8. Traditional error rate in terms of application category

***Source:*** *Smart City Infrastructure from Bashundhara Shopping Complex, Bangladesh; Healthcare Monitoring from Asgar Ali Hospital and Anwar Khan Modern Hospital, Bangladesh; Industrial IoT from Pran Apparels, Bangladesh, 2024.*

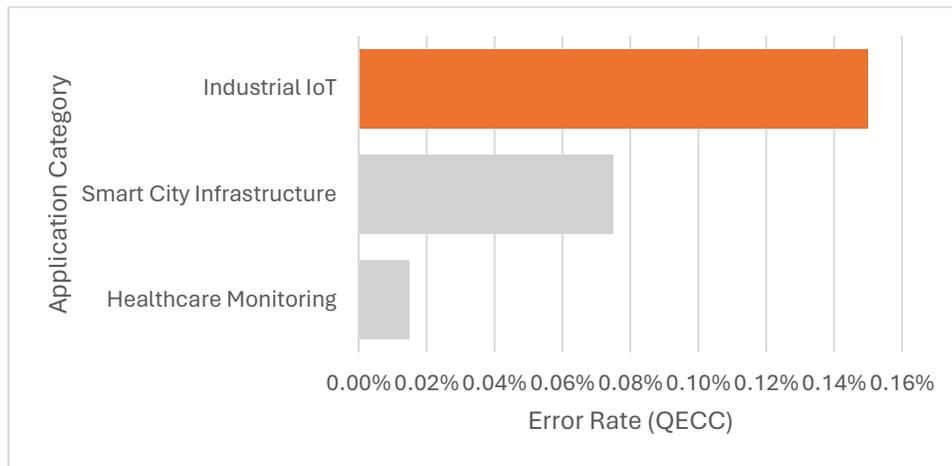

Figure 9. QECC error rate in terms of application category

***Source:*** *Smart City Infrastructure from Bashundhara Shopping Complex, Bangladesh; Healthcare Monitoring from Asgar Ali Hospital and Anwar Khan Modern Hospital, Bangladesh; Industrial IoT from Pran Apparels, Bangladesh, 2024.*

Figure 10 shows a positive correlation between error rates of traditional computing and Quantum-Edge Cloud Computing (QECC). QECC has a lower error rate, but its performance is proportional to that of traditional computing. Factors causing errors could affect both traditional and QECC systems. The graph's tight cluster of data points suggests that improvements or deteriorations in traditional systems could be a predictor of similar trends in QECC systems.

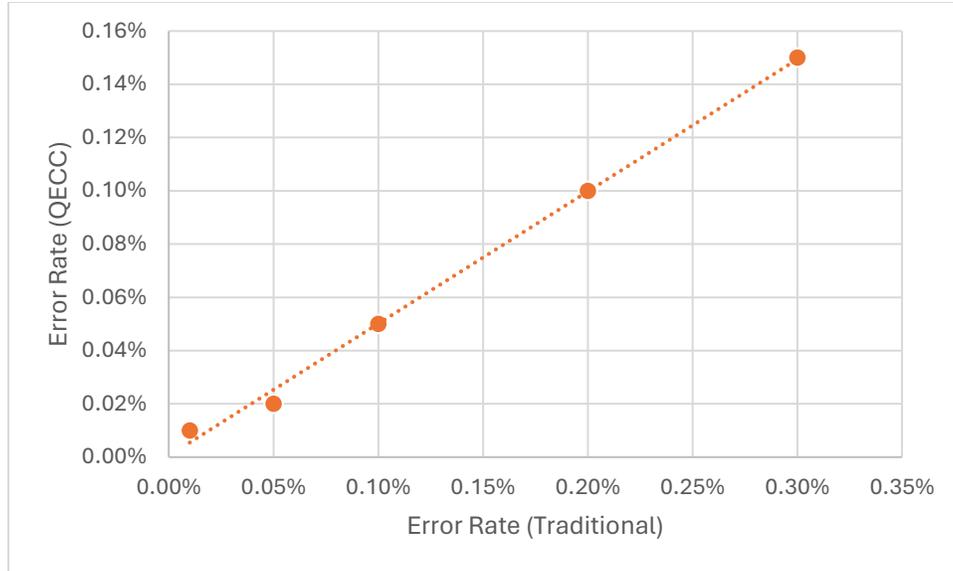

Figure 10. Correlation between Traditional and QECC Error Rate

***Source:*** *Smart City Infrastructure from Bashundhara Shopping Complex, Bangladesh; Healthcare Monitoring from Asgar Ali Hospital and Anwar Khan Modern Hospital, Bangladesh; Industrial IoT from Pran Apparels, Bangladesh, 2024.*

**4.1. Case Study: Enhancing Autonomous Vehicle Safety and Accuracy with Quantum Sensors**

Autonomous vehicles represent a significant leap forward in automotive technology, offering the promise of safer, more efficient roads. However, the reliability of these vehicles heavily depends on the accuracy and sensitivity of their onboard sensors for object detection, navigation, and overall situational awareness. A primary challenge in the development of autonomous vehicles is ensuring the precision of object detection under varying environmental conditions. Traditional sensor technologies like LiDAR, though widely used, encounter limitations in accuracy, especially in adverse weather conditions such as fog or heavy rain. These limitations can potentially compromise safety and reliability.

*Current Sensor Technology:* **LiDAR (Light Detection and Ranging):** LiDAR sensors play a crucial role in creating a 3D map of the vehicle's surroundings by emitting light pulses and measuring the reflections. Despite their capabilities, LiDAR systems are not only costly but also face challenges in maintaining accuracy during poor weather conditions.

*Quantum Solution:* Quantum sensors emerge as a groundbreaking solution, leveraging the principles of quantum mechanics to surpass the sensitivity and accuracy limits of classical sensors. These advanced sensors can detect smaller, fainter signals, offering a significantly refined perception of the environment.

*Benefits Achieved:*
- **Improved Object Detection:** Quantum sensors enhance the vehicle's ability to precisely identify various objects such as pedestrians, cyclists, and other vehicles, even in challenging weather scenarios.
- **Enhanced Safety:** The accuracy provided by quantum sensors allows for better decision-making and quicker responses to potential hazards, thus improving the overall safety of autonomous vehicles.

- **Reduced Reliance on GPS:** Quantum sensors offer a dependable alternative for navigation in situations where GPS signals are weak or entirely absent.

*Current Stage:* Although still in the development phase, the potential of quantum sensors in autonomous vehicles is being actively explored. For instance, Hyundai has initiated collaborations with quantum computing companies to investigate the application of quantum algorithms in processing LiDAR data for enhanced object detection capabilities.

*Future Outlook:* The integration of quantum sensors with existing technologies like LiDAR presents a promising avenue for advancements in autonomous vehicle safety, reliability, and performance in all types of weather conditions. As technology matures and becomes more economically feasible, quantum sensors are poised to play a pivotal role in the evolution of autonomous driving systems.

## 5. Discussion

The empirical data presented in the study demonstrates that Quantum Edge Cloud Computing (QECC) can significantly enhance processing speeds and reduce error rates in Internet of Things (IoT) applications when compared to traditional computing. The improvement is particularly noteworthy in Industrial IoT, which aligns with the theoretical predictions of quantum computing's impact on computational efficiency. Moreover, the integration of edge computing has shown a positive correlation with the reduction in latency, validating the literature's hypotheses regarding the benefits of localized data processing. However, the error rates associated with QECC indicate that challenges still exist in optimizing these systems for error resilience, particularly at higher processing speeds and throughputs. The case studies presented by various sectors in Bangladesh have shed light on the real-world impact of QECC on smart city infrastructure and healthcare monitoring, presenting a promising outlook for its application in diverse environments. Nevertheless, these case studies also highlight the necessity for further research, especially in scaling these solutions and addressing the practical challenges of integration across varied IoT platforms. The findings underscore the potential of QECC but also caution against a one-size-fits-all approach, emphasizing the need for tailored solutions that consider each application domain's unique demands and technical capabilities. As the study suggests, future research directions must include the development of quantum-resistant encryption to safeguard against emerging security threats and optimizing quantum algorithms specifically for edge computing environments. The research outcomes with theoretical insights, offering a critical analysis of QECC's current state and charting a path forward for realizing its full potential in the IoT landscape.

## 6. Recommendations and Future Research Directions

Based on the study's findings and in conjunction with the literature review and case studies examined, the following recommendations and future research directions can be proposed.

**Recommendations:**
1. Optimization of QECC for Industrial IoT: Given the higher error rates in traditional computing identified in Industrial IoT applications, focused efforts on optimizing QECC for this sector should be a priority. This includes enhancing data integrity and processing capabilities tailored to the heavy-industrial context.

2. Expansion of QECC in Healthcare and Smart Cities: The study shows promising results in healthcare monitoring and smart city infrastructure. Initiatives should be expanded to leverage QECC for broader applications in these domains to further improve efficiency and response times.
3. Improvement in Quantum Algorithm Efficiency: With the observed limitations at higher data throughputs, it is recommended to refine quantum algorithms to ensure they maintain low error rates even as processing speeds and throughputs increase.

**Future Research Directions:**
1. Development of Quantum-Resistant Cryptography: As quantum computing evolves, so do potential security threats. Future studies should investigate creating encryption methods that are resistant to quantum decryption techniques.
2. Diversification of Case Studies: To understand the broader implications of QECC across various geographies and sectors, future research should include a more diverse range of case studies, including those outside Bangladesh.
3. Interoperability and Standardization: As QECC becomes more prevalent, establishing interoperability standards will be critical. Future research should address the seamless integration of quantum, edge, and cloud computing within existing IoT ecosystems.
4. Energy Efficiency: Given the energy demands of computing, especially quantum computing, future research should also focus on making QECC more energy-efficient and environmentally sustainable.

By adhering to these recommendations and focusing on the outlined research directions, the potential of QECC to transform IoT applications can be fully realized, leading to a new standard of performance, security, and scalability in the IoT domain.

**7. Limitations of this Study**

The study on Quantum-Edge Cloud Computing (QECC) for IoT applications has some limitations. It relies on data from only a few sectors in Bangladesh, and quantum computing technology is still in the early development stages. The case studies selected represent only a narrow slice of potential uses, and the integration of quantum, edge, and cloud computing poses complex challenges. IoT systems are commonly constructed using conventional computing methods, potentially lacking compatibility with quantum computing. Consequently, organizations must allocate resources towards acquiring new infrastructure and expertise to seamlessly integrate quantum technologies into their current IoT systems. Quantum hardware is another limitation. Quantum processors are currently in their initial phases, with only a handful of companies providing access to such computing resources. This scarcity poses a challenge for organizations aiming to test and expand quantum computing applications. Resource constraints and evolving cyber threats are also significant barriers. Finally, the study suggests optimizing quantum algorithms for edge computing, but practical techniques are not substantiated. Further research is necessary to overcome these obstacles and fully realize QECC's potential in the IoT sector.

**8. Conclusions**

This study has explored the potential of Quantum-Edge Cloud Computing (QECC) to transform the Internet of Things (IoT) by leveraging the unique strengths of quantum computing, edge computing, and cloud computing. Our investigation into the theoretical foundations and real-world applications reveals QECC as a robust solution for enhancing data processing speeds, security, and scalability across IoT systems. Key findings emphasize the critical need for developing quantum-resistant encryption methods to protect against future quantum threats, alongside addressing scalability and standardization challenges to ensure widespread adoption. Future research focusing on optimizing quantum algorithms for edge environments and enhancing interoperability through standardized protocols will be pivotal. As QECC continues to evolve, it promises to overcome current IoT limitations, heralding a new era of technological innovation that is both secure and efficient, driven by interdisciplinary collaboration and continuous advancements.